\documentclass[prb,twocolumn,showpacs,preprintnumbers,amsmath,amssymb]{revtex4}

\usepackage{graphicx}
\begin{document}


\title{ Nodal Superconductivity with Multiple Gaps in SmFeAsO$_{0.9}$F$_{0.1}$ }

\author{Yong-Lei Wang}
\author{Lei Shan}\email{lshan@aphy.iphy.ac.cn}
\author{Lei Fang, Peng Cheng, Cong Ren}
\author{Hai-Hu Wen}\email{hhwen@aphy.iphy.ac.cn}

\affiliation{National Laboratory for Superconductivity, Institute of Physics $\&$ Beijing National
Laboratory for Condensed Matter Physics, Chinese Academy of Sciences, P.O. Box 603, Beijing 100190,
China}

\date{\today}

\begin{abstract}
We report the observation of two gaps in the superconductor SmFeAsO$_{0.9}$F$_{0.1}$ (F-SmFeAsO)
with $T_c=51.5K$ as measured by point-contact spectroscopy. Both gaps decrease with temperature and
vanish at $T_c$ and the temperature dependence of the gaps are described by the theoretical
prediction of the Bardeen-Cooper-Schrieffer (BCS) theory. A zero-bias conductance peak (ZBCP) was
observed, indicating the presence of Andreev bound states at the surface of F-SmFeAsO. Our results
strongly suggest an unconventional nodal superconductivity with multiple gaps in F-SmFeAsO.

\end{abstract}


\maketitle


\section{introduction}

Recently, superconductivity was discovered in Fe-based layered superconductor
LaFeAs[O$_{1-x}$F$_x$] (F-LaFeAsO) \cite{discovery_Kamihara2008JACS}. By replacing La by some other
rare earth elements, $T_c$ was improved quickly up to $55$K \cite{F-SmOFeAs_RenZA2008cond} which is
substantially higher than that of the single-layered cuprate superconductors and exceeds the
theoretical value predicted by the conventional BCS theory \cite{Tc_McMillan1968pr}. The
superconductivity was also observed in hole doped case
\cite{Hole_WenHH2008EPL,Hole_RotterM2008cond}. Thus a new family of high-$T_c$ superconductors was
opened up, providing an unique chance to understand high-$T_c$ superconductivity. To detect the gap
structure and pairing symmetry is an essential step to reveal the mechanism of these Fe-based
superconductors. The specific heat measurement on F-LaFeAsO showed a nonlinear magnetic field
dependence of the electronic specific heat coefficient as expected by a nodal superconductor
\cite{SpecificHeat_MuG2008cond}. This was proved subsequently by the measurements of point-contact
spectroscopy \cite{PCT-LaOFFeAs_ShanL2008cond}, lower critical field $H_{c1}$
\cite{Hc1_RenC2008cond}, London penetration depth $\lambda$ and spin-lattice relaxation rate
$1/T_1$ \cite{NMR_AhilanK2008cond,NMR_NakaiY2008cond}. The muon spin relaxation measurements also
presented the possibility of dirty $d$-wave pairing \cite{MuSR_LuetkensH2008cond}. Theoretically,
some calculations support $d$-wave
\cite{TwobandSc_YaoZJ2008cond,d-wave_LiT2008cond,d-wave_SiQM2008cond} while others support an
unconventional $s$-wave pairing \cite{s-wave_MazinII2008cond,s-wave_KurokiK2008cond}. Most
surprisingly, recent Andreev reflection data favor an isotropic single gap in
SmFeAsO$_{0.85}$F$_{0.15}$ and NdFeAsO$_{0.85}$ \cite{PCT_ChenTY2008cond,PCT_YatesKA2008cond},
while a nodal supercoductivity with two-gap structure was suggested by the NMR experiment on
PrFeAsO$_{0.89}$F$_{0.11}$ \cite{NMR_ZhengGQ2008cond}. Actually, multi-band superconductivity has
been predicted theoretically \cite{TwobandSc_LiJ2008CPL,
TwobandSc_Marsiglio2008cond,TwobandSc_YaoZJ2008cond,TwobandSc_WangZH2008cond,TwobandSc_HanQ2008EPL}
and was supported by the magnetic properties of SmFeAsO$_{0.8}$F$_{0.2}$ single crystal
\cite{MagAnis_WeyenethS2008cond} and the high-magnetic field resistance of
LaFeAsO$_{0.89}$F$_{0.11}$ \cite{HighMag_HunteF2008Nature}. However, multiple gaps have not been
detected for the moment by both Andreev reflection or tunneling experiment which is a powerful tool
to measure the superconducting gap. Therefore, more experiments on the samples with higher quality
are strongly desired.

In this paper, we present the point-contact spectroscopy data of the compact and rigid
SmFeAsO$_{0.9}$F$_{0.1}$ (F-SmFeAsO) samples. A ZBCP was observed repeatedly, indicating the nodal
gap structure of F-SmFeAsO similar to that of F-LaFeAsO. Moreover, two different gaps were observed
and both of them decrease with temperature and vanish at $T_c$. These results strongly suggest that
F-SmFeAsO has an unconventional nodal superconductivity with multiple gaps.

\begin{figure}[]
\includegraphics[scale=1.3]{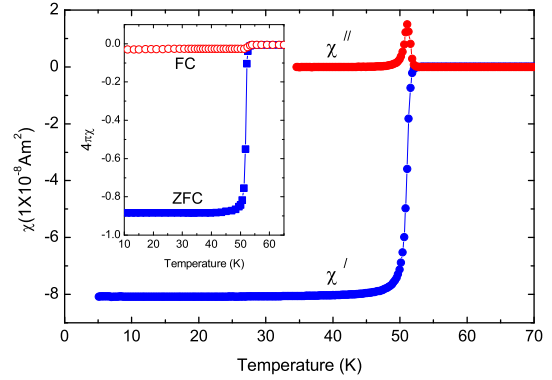}
\caption{\label{fig:fig1} (Color online) Temperature dependence of AC susceptibility of F-SmFeAsO.
Inset: DC susceptibility of F-SmFeAsO. }
\end{figure}

\section{experiment}

The superconducting F-SmFeAsO samples were prepared by a high pressure synthesis method
\cite{HP_RenZA2008cond}. The detailed information about the synthesization is elaborated in a
recent paper \cite{F-SmOFeAs_RenZA2008cond}. As shown in Fig.~\ref{fig:fig1}, DC susceptibility
(measured under a magnetic field of 1 Oe) and AC susceptibility data (measured using an AC
amplitude of 0.1 Oe) exhibit a sharp magnetic transition. The width defined between the 10\% and
90\% cuts of the transition is below 2K, with the middle of the Meissner transition at 51.5 K,
indicating the good quality of the superconducting phase. Compared with the samples synthesized by
the common vacuum quartz tube synthesis method, the samples studied here are much more compact and
rigid, thus more suitable for point-contact spectroscopy measurements. The point-contact junctions
are prepared by carefully driving the Pt/Ir alloy or Au tips towards the sample surface which is
polished and cleaned beforehand. The tip's preparation and the details of the experimental setup
were described elsewhere \cite{tip-prepare_ShanL2003PRB}. Typical four-terminal and lock-in
techniques were used to measure the conductance-voltage ($dI/dV-V$ or $G-V$) characteristics. Each
measurement is comprised of two successive cycles, to check the absence of heating-hysteresis
effects.

\begin{figure}[]
\includegraphics[scale=1.2]{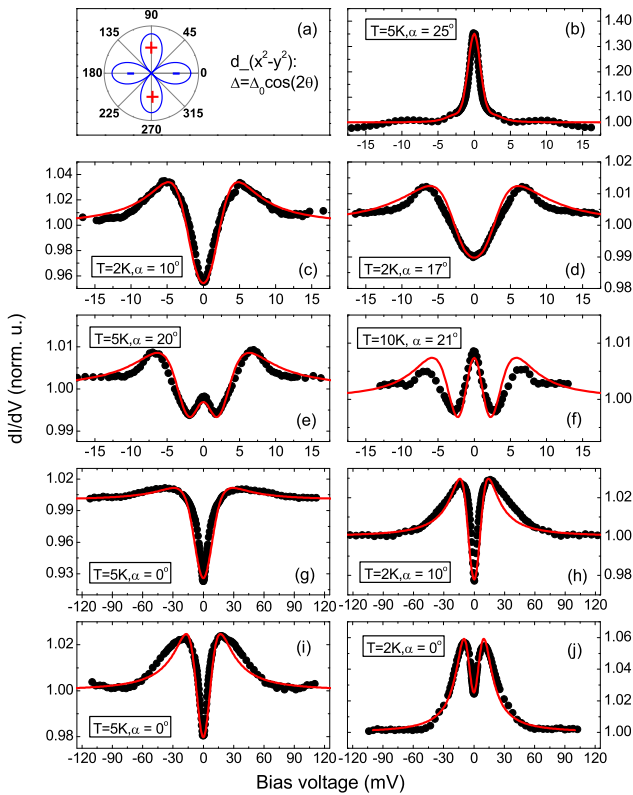}
\caption{\label{fig:fig2} (Color online) (a) Gap structure (d-wave type) used to fit the
experimental data. (b)-(j) Experimental data (solid circles) measured at various locations on the
sample surface and theoretical calculations (solid lines) with the gap function shown in (a). }
\end{figure}

\section{results and discussions}

Figure~\ref{fig:fig2} shows the $G-V$ curves measured at various locations on the sample surface,
which have been normalized by the high-bias data. A distinct ZBCP can be seen in
Figs.~\ref{fig:fig2}(b),(e) and (f), indicating the existence of surface Andreev bound states,
which is a clear signature of the pair potentials with reversal sign in momentum space and is known
as an unique character of nodal superconductors \cite{Deutscher2005RMP}. Such ZBCP has also been
observed in F-LaFeAsO \cite{PCT-LaOFFeAs_ShanL2008cond} and F-NdFeAsO \cite{PCAR_Samuely2008cond},
thus nodal superconductivity is most possibly a common property of the Fe-based superconductors.
Blonder {\it et al.} \cite{BTK_BlonderGE1982PRB} have proposed a simplified theory for the $G-V$
curves of an $s$-wave superconductor/normal metal junction separated by a barrier of arbitrary
strength. The barrier strength is parameterized by a dimensionless number $Z$ which describes the
crossover from metallic to ideal tunnel junction behavior by $Z=0$ to $Z=\infty$. Obviously, this
$s$-wave model can not explain the observed ZBCP. Tanaka and Kashiwaya
\cite{ZBCP-KashiwayaS2000RPP} extended the BTK model to deal with the issue of unconventional
pairing symmetry. In this case, the angle ($\alpha$) between the quasiparticle injecting direction
and the main crystalline axis was introduced as another parameter. Moreover, the isotropic
superconducting gap $\Delta$ was replaced by an anisotropic gap with $d_{x^2-y^2}$ symmetry:
$\Delta=\Delta_0cos(2\theta)$, as shown in Fig.~\ref{fig:fig2}(a). It was then predicted that, for
$Z>0$ the ZBCP is formed for all directions in the $a-b$ plane except when tunneling into the (100)
and (010) planes. This ZBCP will be suppressed when the quasiparticle scattering or surface
roughness is strong enough \cite{ZBCP-ApriliM1998PRBr,ZBCP_Tanaka2002PRB}. It was found that all
the spectra shown in Fig.~\ref{fig:fig2} can be described very well by this extended BTK model.
Moreover, various quasiparticle injecting angles ($\alpha$) are obtained for the spectra measured
at different positions, indicating that our measurement is a local detection. The broadening
parameter $\Gamma/\Delta_0$ \cite{Broad_PlecenA1994PRB} in the calculations is between 0.2 and 1,
coming from some unclear scattering mechanisms at the interface, which is similar to the case of
cuprate superconductors. It should be mentioned that the gap structure with reversal sign is
necessary to explain our data while the details of the gap function can not be distinguished
although the $d_{x^2-y^2}$-wave symmetry was accepted here (for example, $d_{xy}$-wave symmetry
\cite{dxy_Sachdev2002PhysicaA} can explain our data as well).

Another remarkable find in the calculations presented in Fig.~\ref{fig:fig2} is that, the
determined maximum gap $\Delta_0$ can be divided into two groups, namely, a big gap of $10.5\pm
0.5$ meV and a small gap of $3.7\pm 0.4$ meV, though the barrier strength $Z$ and quasiparticle
injecting angle $\alpha$ are random due to the diverse configuration of the grains in the
polycrystalline samples. The small gap can not be explained as degradation of the sample surface
since its value is distributed in a narrow range far below that of the big gap. To our knowledge,
this multi-gap feature is observed for the first time in transition metal-based high-$T_c$
superconductors. To get further insight into this point, we have measured the temperature
dependence of these two gaps with distinct energy scales.

\begin{figure}
\includegraphics[scale=1.4]{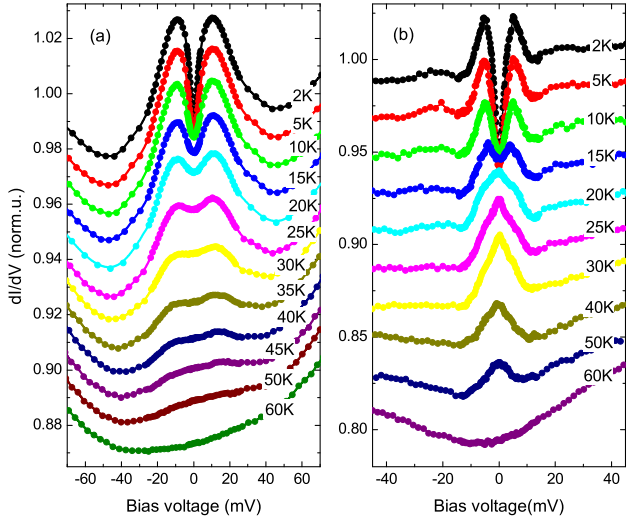}
\caption{\label{fig:fig3} (Color online) Conductance measured for various temperatures from 2K to
above $T_c$. (a) large gap, (b) small gap. All the curves except the top one are offset downwards
for clarity. }
\end{figure}

\begin{figure}
\includegraphics[scale=1.2]{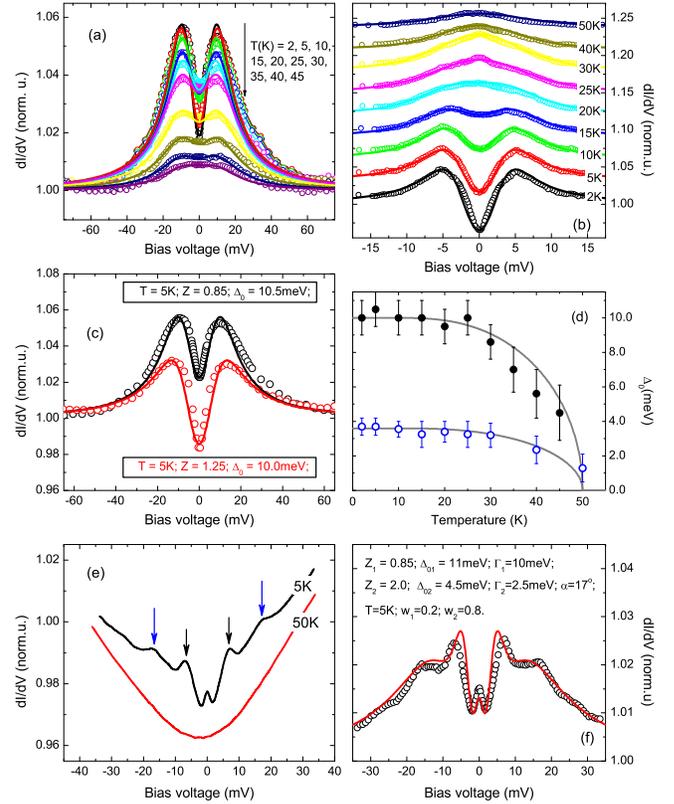}
\caption{\label{fig:fig4} (Color online) (a) and (b) Temperature dependence of normalized spectra
corresponding to Figs.~\ref{fig:fig2}(a) and (b), respectively. The solid lines are theoretical
calculations according to the extended BTK model. (c) Two spectra measured at same position while
with different junction resistance, solid lines are theoretical calculations. (d) Temperature
dependencies of the gap values determined from fits as shown in (a) and (b), solid lines are the
prediction of BCS theory. (e) The spectra measured at low temperature and around $T_c$ on same
position of the sample surface. (f) The low-temperature spectra after normalization according to
the high-temperature one as shown in (e), the solid line is the theoretical fit.}
\end{figure}

Figure~\ref{fig:fig3} shows the temperature dependence of two types of spectra corresponding two
different gaps as mentioned above. At lower temperatures, two coherence peaks can be seen clearly
accompanied by low-energy depression of the quasiparticle density of states. With increasing
temperature, the peaks are suppressed and smeared continuously and finally the spectra become a
smooth featureless curve around $T_c$. These data were normalized (by the data of $T=60$K) in
order to be compared with theoretical models, as shown in Figs.~\ref{fig:fig4}(a) and (b). It was
found that all the data can be fitted very well to the extended BTK model mentioned above.
Fig.~\ref{fig:fig4}(c) shows two spectra measured at the same location while with different
junction resistances and hence different barrier strengths. The similar gap value about 10 meV was
obtained from these spectra, indicating that the detected superconductivity does not depend on the
junction resistance. Fig.~\ref{fig:fig4}(d) summarizes the gap values obtained from
Figs.~\ref{fig:fig4}(a) and (b). The temperature dependence of both the large gap and the small
one can be described by the prediction of BCS theory. The gap value of $\Delta_0=$10meV leads to
$\Delta_0/k_BT_c=2.3$, a bit larger than the prediction of weak-coupling $d$-wave BCS theory.
Moreover, both gaps are closed around $T_c$ reflecting the inter-band coupling existing in this
material.

It is interesting to note that some spectra exhibit a two-gap feature similar to that of MgB$_2$.
As shown in Fig.~\ref{fig:fig4}(e), compared with the high-temperature spectrum measured around
$T_c$, the low-temperature one has three distinct features: a ZBCP, two coherence peaks, and two
symmetric hump at higher energy. In Fig.~\ref{fig:fig4}(f), we try to simulate the low-temperature
spectrum after normalization according to the high-temperature one. In the calculation, a simple
two-component BTK model was accepted in which the normalized conductance ($G=dI/dV$) is expressed
by $G=w_1G_1+w_2G_2$, where $G_1$ and $G_2$ are the conductance associated with the large gap and
small gap, respectively, $w_1$ and $w_2$ ($w_1+w_2=1$) are corresponding weights of these two gaps
contributing to the total conductance. In order to reduce the number of fitting parameters, the
$s$-wave symmetry was assumed for the large gap while the $d_{x^2-y^2}$-wave symmetry for the
small gap. It was found that the main features in the spectrum can be fitted very well. Although
this is a rough approximation, it captures the main physics of multiple gaps and nodal
superconductivity. Most interestingly, the determined gap values from the calculation are 11meV
and 4.5meV, very close to the results obtained from the spectra measured at other locations.

Besides the sample of F-SmFeAsO which is focused on in this work, we have also measured the
point-contact spectra of F-LaFeAsO \cite{PCT-LaOFFeAs_ShanL2008cond} and F-NdFeAsO. In those
samples, a more prominent ZBCP was often observed while some confused backgrounds can also be
observed occasionally compared with the data of F-SmFeAsO, thus quantitative analysis is more
difficult. A most possible explanation for this difference is that, F-SmFeAsO studied here was
prepared by high-pressure technique while other samples were synthesized in atmospheric pressure
and hence more fragile and loose. Therefore, there is often a space between adjacent grains, which
can be seen clearly in the SEM photographs. Consequently, the tip is easy to penetrate through the
sample surface and rests on a pit, leading to multiple contacts between the tip and surrounded
micro-crystals. On the one hand, it increases the opportunities to detect strong ZBCP if a nodal
pairing symmetry exists. On the other hand, it will induce a more confused background due to the
complicated interface effect. In this case, the gap values can still be estimated by using a
convenient method prosed by Dagan {\it et al.} \cite{Gap-DaganY2000PRB}. The method of data
analysis consists simply in subtracting conductances measured in an applied field from the zero
field conductance. Since the maximum of the density of states at the gap value should be sensitive
to an applied field, such subtraction will leads to a dip around the gap value in the subtracted
spectra \cite{Gap-DaganY2000PRB}. Using this method, we have estimated $\Delta_0\approx3.9$meV for
F-LaFeAsO, which is consistent with the results from both specific heat
\cite{SpecificHeat_MuG2008cond} and lower critical field \cite{Hc1_RenC2008cond}.

\begin{figure}[]
\includegraphics[scale=1.4]{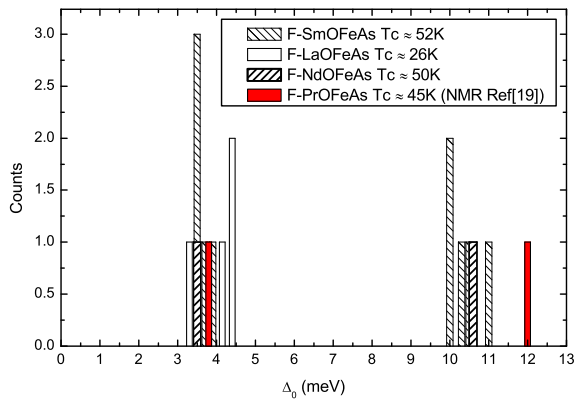}
\caption{\label{fig:fig5} (Color online) Statistical chart of the $\Delta_0$-values determined from
the spectra measured at various positions on the sample surfaces. The vertical axis denotes the
occurring times for a given gap-value specified by the horizontal-axis. }
\end{figure}

In Fig.~\ref{fig:fig5}, we sum up the gap values of some Fe-based superconductors with different
$T_c$, which were obtained from our point-contact spectroscopy measurements. The result of
PrFeAsO$_{0.89}$F$_{0.11}$ from NMR experiment \cite{NMR_ZhengGQ2008cond} is also presented for
comparison. It was noted that for all these Fe-based superconductors, the superconducting gaps can
be divided into two groups with distinct energy scales centered at 4meV and 11meV, respectively.
Recently, the Andreev reflection data suggested a gap of $\Delta\approx6.7$meV for F-SmFeAsO with
$T_c=42$K and $\Delta\approx7$meV with $T_c=45.5$K. However, the isotropic $s$-wave gap was
accepted in these analysis. If a $d$-wave gap was assumed, the obtained maximum gap $\Delta_0$ is
about 8meV, which should be ascribed to the big gap obtained in this work. Therefore, the high
$T_c$ can be achieved now is most possibly dominated by the big gap. However, the detailed gap
structure is still an open issue until the high-quality single crystals can be obtained.

\section{summary}
In summary, we have studied point-contact spectroscopy of the junctions built up between a
normal-metal tip and the newly discovered Fe-based layered superconductor SmFeAsO$_{0.9}$F$_{0.1}$.
A zero-bias conductance peak was observed and demonstrated to be related to the surface Andreev
bound states. Two superconducting gaps with different energy scales were observed which depend on
the temperature in the similar way as predicted by BCS theory. Our data present strong evidence
that SmFeAsO$_{0.9}$F$_{0.1}$ is a nodal superconductor with multiple gaps.

Note added: Just before submitting this manuscript we have noticed a scanning tunneling
spectroscopy study on SmFeAsO$_{0.85}$\cite{STS_Millo2008cond} which prepared by the same group
using the same method as that in this work. The obtained spectra are in good agreement with our
observations in Fig.~\ref{fig:fig2}(c) and (f).

\begin{acknowledgments}
We are grateful to Prof. Zhongxian Zhao and Dr. Zhian Ren for providing us the high quality
SmFeAsO$_{0.9}$F$_{0.1}$ samples made by high pressure technique. This work is supported by the
Natural Science Foundation of China, the Ministry of Science and Technology of China ( 973 project
No: 2006CB601000, 2006CB921802, 2006CB921300 ), and Chinese Academy of Sciences (Project ITSNEM).
\end{acknowledgments}

\bibliography{FSmFeAsO}

\begin{thebibliography}{37}
\expandafter\ifx\csname natexlab\endcsname\relax\def\natexlab#1{#1}\fi
\expandafter\ifx\csname bibnamefont\endcsname\relax
  \def\bibnamefont#1{#1}\fi
\expandafter\ifx\csname bibfnamefont\endcsname\relax
  \def\bibfnamefont#1{#1}\fi
\expandafter\ifx\csname citenamefont\endcsname\relax
  \def\citenamefont#1{#1}\fi
\expandafter\ifx\csname url\endcsname\relax
  \def\url#1{\texttt{#1}}\fi
\expandafter\ifx\csname urlprefix\endcsname\relax\def\urlprefix{URL }\fi
\providecommand{\bibinfo}[2]{#2}
\providecommand{\eprint}[2][]{\url{#2}}

\bibitem[{\citenamefont{Kamihara et~al.}(2008)\citenamefont{Kamihara, Watanabe,
  Hirano, and Hosono}}]{discovery_Kamihara2008JACS}
\bibinfo{author}{\bibfnamefont{Y.}~\bibnamefont{Kamihara}},
  \bibinfo{author}{\bibfnamefont{T.}~\bibnamefont{Watanabe}},
  \bibinfo{author}{\bibfnamefont{M.}~\bibnamefont{Hirano}}, \bibnamefont{and}
  \bibinfo{author}{\bibfnamefont{H.}~\bibnamefont{Hosono}},
  \bibinfo{journal}{J.\ Am.\ Chem.\ Soc.} \textbf{\bibinfo{volume}{130}},
  \bibinfo{pages}{3296} (\bibinfo{year}{2008}).

\bibitem[{\citenamefont{Ren et~al.}(2008{\natexlab{a}})\citenamefont{Ren, Lu,
  Yang, Yi, Shen, Li, Che, Dong, Sun, Zhou et~al.}}]{F-SmOFeAs_RenZA2008cond}
\bibinfo{author}{\bibfnamefont{Z.~A.} \bibnamefont{Ren}},
  \bibinfo{author}{\bibfnamefont{W.}~\bibnamefont{Lu}},
  \bibinfo{author}{\bibfnamefont{J.}~\bibnamefont{Yang}},
  \bibinfo{author}{\bibfnamefont{W.}~\bibnamefont{Yi}},
  \bibinfo{author}{\bibfnamefont{X.~L.} \bibnamefont{Shen}},
  \bibinfo{author}{\bibfnamefont{Z.~C.} \bibnamefont{Li}},
  \bibinfo{author}{\bibfnamefont{G.~C.} \bibnamefont{Che}},
  \bibinfo{author}{\bibfnamefont{X.~L.} \bibnamefont{Dong}},
  \bibinfo{author}{\bibfnamefont{L.~L.} \bibnamefont{Sun}},
  \bibinfo{author}{\bibfnamefont{F.}~\bibnamefont{Zhou}}, \bibnamefont{et~al.},
  \bibinfo{journal}{Chin.\ Phys.\ Lett.} \textbf{\bibinfo{volume}{25}},
  \bibinfo{pages}{2215} (\bibinfo{year}{2008}{\natexlab{a}}).

\bibitem[{\citenamefont{McMillan}(1968)}]{Tc_McMillan1968pr}
\bibinfo{author}{\bibfnamefont{W.~L.} \bibnamefont{McMillan}},
  \bibinfo{journal}{Phys.\ Rev.} \textbf{\bibinfo{volume}{167}},
  \bibinfo{pages}{331} (\bibinfo{year}{1968}).

\bibitem[{\citenamefont{Wen et~al.}(2008)\citenamefont{Wen, Mu, Fang, Yang, and
  Zhu}}]{Hole_WenHH2008EPL}
\bibinfo{author}{\bibfnamefont{H.-H.} \bibnamefont{Wen}},
  \bibinfo{author}{\bibfnamefont{G.}~\bibnamefont{Mu}},
  \bibinfo{author}{\bibfnamefont{L.}~\bibnamefont{Fang}},
  \bibinfo{author}{\bibfnamefont{H.}~\bibnamefont{Yang}}, \bibnamefont{and}
  \bibinfo{author}{\bibfnamefont{X.}~\bibnamefont{Zhu}},
  \bibinfo{journal}{EuroPhys.\ Lett.} \textbf{\bibinfo{volume}{82}}
  (\bibinfo{year}{2008}).

\bibitem[{\citenamefont{Rotter et~al.}()\citenamefont{Rotter, Tegel, and
  Johrendt}}]{Hole_RotterM2008cond}
\bibinfo{author}{\bibfnamefont{M.}~\bibnamefont{Rotter}},
  \bibinfo{author}{\bibfnamefont{M.}~\bibnamefont{Tegel}}, \bibnamefont{and}
  \bibinfo{author}{\bibfnamefont{D.}~\bibnamefont{Johrendt}},
  \bibinfo{howpublished}{arXiv:0805.4630}.

\bibitem[{\citenamefont{Mu et~al.}(2008)\citenamefont{Mu, Zhu, Fang, Shan, Ren,
  and Wen}}]{SpecificHeat_MuG2008cond}
\bibinfo{author}{\bibfnamefont{G.}~\bibnamefont{Mu}},
  \bibinfo{author}{\bibfnamefont{X.}~\bibnamefont{Zhu}},
  \bibinfo{author}{\bibfnamefont{L.}~\bibnamefont{Fang}},
  \bibinfo{author}{\bibfnamefont{L.}~\bibnamefont{Shan}},
  \bibinfo{author}{\bibfnamefont{C.}~\bibnamefont{Ren}}, \bibnamefont{and}
  \bibinfo{author}{\bibfnamefont{H.-H.} \bibnamefont{Wen}},
  \bibinfo{journal}{Chin.\ Phys.\ Lett.} \textbf{\bibinfo{volume}{25}},
  \bibinfo{pages}{2221} (\bibinfo{year}{2008}).

\bibitem[{\citenamefont{Shan et~al.}()\citenamefont{Shan, Wang, Zhu, Mu, Fang,
  and Wen}}]{PCT-LaOFFeAs_ShanL2008cond}
\bibinfo{author}{\bibfnamefont{L.}~\bibnamefont{Shan}},
  \bibinfo{author}{\bibfnamefont{Y.}~\bibnamefont{Wang}},
  \bibinfo{author}{\bibfnamefont{X.}~\bibnamefont{Zhu}},
  \bibinfo{author}{\bibfnamefont{G.}~\bibnamefont{Mu}},
  \bibinfo{author}{\bibfnamefont{L.}~\bibnamefont{Fang}}, \bibnamefont{and}
  \bibinfo{author}{\bibfnamefont{H.-H.} \bibnamefont{Wen}},
  \bibinfo{howpublished}{ArXiv:0803.2405 [cond-mat.supr-con]}.

\bibitem[{\citenamefont{Ren et~al.}()\citenamefont{Ren, Wang, Yang, Zhu, Fang,
  Mu, Shan, and Wen}}]{Hc1_RenC2008cond}
\bibinfo{author}{\bibfnamefont{C.}~\bibnamefont{Ren}},
  \bibinfo{author}{\bibfnamefont{Z.-S.} \bibnamefont{Wang}},
  \bibinfo{author}{\bibfnamefont{H.}~\bibnamefont{Yang}},
  \bibinfo{author}{\bibfnamefont{X.}~\bibnamefont{Zhu}},
  \bibinfo{author}{\bibfnamefont{L.}~\bibnamefont{Fang}},
  \bibinfo{author}{\bibfnamefont{G.}~\bibnamefont{Mu}},
  \bibinfo{author}{\bibfnamefont{L.}~\bibnamefont{Shan}}, \bibnamefont{and}
  \bibinfo{author}{\bibfnamefont{H.-H.} \bibnamefont{Wen}},
  \bibinfo{howpublished}{ArXiv:0804.1726 [cond-mat.supr-con]}.

\bibitem[{\citenamefont{Ahilan et~al.}()\citenamefont{Ahilan, Ning, Imai,
  Sefat, Jin, McGuire, Sales, and Mandrus}}]{NMR_AhilanK2008cond}
\bibinfo{author}{\bibfnamefont{K.}~\bibnamefont{Ahilan}},
  \bibinfo{author}{\bibfnamefont{F.~L.} \bibnamefont{Ning}},
  \bibinfo{author}{\bibfnamefont{T.}~\bibnamefont{Imai}},
  \bibinfo{author}{\bibfnamefont{A.~S.} \bibnamefont{Sefat}},
  \bibinfo{author}{\bibfnamefont{R.}~\bibnamefont{Jin}},
  \bibinfo{author}{\bibfnamefont{M.~A.} \bibnamefont{McGuire}},
  \bibinfo{author}{\bibfnamefont{B.~C.} \bibnamefont{Sales}}, \bibnamefont{and}
  \bibinfo{author}{\bibfnamefont{D.}~\bibnamefont{Mandrus}},
  \bibinfo{howpublished}{arXiv:0804.4026 [cond-mat.supr-con]}.

\bibitem[{\citenamefont{Nakai et~al.}(2008)\citenamefont{Nakai, Ishida,
  Kamihara, Hirano, and Hosono}}]{NMR_NakaiY2008cond}
\bibinfo{author}{\bibfnamefont{Y.}~\bibnamefont{Nakai}},
  \bibinfo{author}{\bibfnamefont{K.}~\bibnamefont{Ishida}},
  \bibinfo{author}{\bibfnamefont{Y.}~\bibnamefont{Kamihara}},
  \bibinfo{author}{\bibfnamefont{M.}~\bibnamefont{Hirano}}, \bibnamefont{and}
  \bibinfo{author}{\bibfnamefont{H.}~\bibnamefont{Hosono}},
  \bibinfo{journal}{JPSJ} \textbf{\bibinfo{volume}{77}},
  \bibinfo{pages}{073701} (\bibinfo{year}{2008}).

\bibitem[{\citenamefont{{H. Luetkens, H. -H. Klauss, R. Khasanov, A. Amato, R.
  Klingeler, I. Hellmann, N. Leps, A. Kondrat, C. Hess, A. K{\"o}hler, G. Behr,
  J. Werner and B. B{\"u}chner}}()}]{MuSR_LuetkensH2008cond}
\bibinfo{author}{\bibnamefont{{H. Luetkens, H. -H. Klauss, R. Khasanov, A.
  Amato, R. Klingeler, I. Hellmann, N. Leps, A. Kondrat, C. Hess, A.
  K{\"o}hler, G. Behr, J. Werner and B. B{\"u}chner}}},
  \bibinfo{howpublished}{arXiv:0804.3115 [cond-mat.supr-con]}.

\bibitem[{\citenamefont{Yao et~al.}()\citenamefont{Yao, Li, and
  Wang}}]{TwobandSc_YaoZJ2008cond}
\bibinfo{author}{\bibfnamefont{Z.-J.} \bibnamefont{Yao}},
  \bibinfo{author}{\bibfnamefont{J.-X.} \bibnamefont{Li}}, \bibnamefont{and}
  \bibinfo{author}{\bibfnamefont{Z.~D.} \bibnamefont{Wang}},
  \bibinfo{howpublished}{arXiv:0804.4166 [cond-mat.supr-con]}.

\bibitem[{\citenamefont{Li}()}]{d-wave_LiT2008cond}
\bibinfo{author}{\bibfnamefont{T.}~\bibnamefont{Li}},
  \bibinfo{howpublished}{arXiv:0804.0536 [cond-mat.supr-con]}.

\bibitem[{\citenamefont{Si and Abrahams}()}]{d-wave_SiQM2008cond}
\bibinfo{author}{\bibfnamefont{Q.}~\bibnamefont{Si}} \bibnamefont{and}
  \bibinfo{author}{\bibfnamefont{E.}~\bibnamefont{Abrahams}},
  \bibinfo{howpublished}{arXiv:0804.2480 [cond-mat.supr-con]}.

\bibitem[{\citenamefont{Mazin et~al.}()\citenamefont{Mazin, Singh, Johannes,
  and Du}}]{s-wave_MazinII2008cond}
\bibinfo{author}{\bibfnamefont{I.~I.} \bibnamefont{Mazin}},
  \bibinfo{author}{\bibfnamefont{D.~J.} \bibnamefont{Singh}},
  \bibinfo{author}{\bibfnamefont{M.~D.} \bibnamefont{Johannes}},
  \bibnamefont{and} \bibinfo{author}{\bibfnamefont{M.~H.} \bibnamefont{Du}},
  \bibinfo{howpublished}{arXiv:0803.2740 [cond-mat.supr-con]}.

\bibitem[{\citenamefont{Kuroki et~al.}()\citenamefont{Kuroki, Onari, Arita,
  Usui, Tanaka, Kontani, and Aoki}}]{s-wave_KurokiK2008cond}
\bibinfo{author}{\bibfnamefont{K.}~\bibnamefont{Kuroki}},
  \bibinfo{author}{\bibfnamefont{S.}~\bibnamefont{Onari}},
  \bibinfo{author}{\bibfnamefont{R.}~\bibnamefont{Arita}},
  \bibinfo{author}{\bibfnamefont{H.}~\bibnamefont{Usui}},
  \bibinfo{author}{\bibfnamefont{Y.}~\bibnamefont{Tanaka}},
  \bibinfo{author}{\bibfnamefont{H.}~\bibnamefont{Kontani}}, \bibnamefont{and}
  \bibinfo{author}{\bibfnamefont{H.}~\bibnamefont{Aoki}},
  \bibinfo{howpublished}{arXiv:0803.3325 [cond-mat.supr-con]}.

\bibitem[{\citenamefont{Chen et~al.}(2008)\citenamefont{Chen, Tesanovic, Liu,
  Chen, and Chien}}]{PCT_ChenTY2008cond}
\bibinfo{author}{\bibfnamefont{T.~Y.} \bibnamefont{Chen}},
  \bibinfo{author}{\bibfnamefont{Z.}~\bibnamefont{Tesanovic}},
  \bibinfo{author}{\bibfnamefont{R.~H.} \bibnamefont{Liu}},
  \bibinfo{author}{\bibfnamefont{X.~H.} \bibnamefont{Chen}}, \bibnamefont{and}
  \bibinfo{author}{\bibfnamefont{C.~L.} \bibnamefont{Chien}},
  \bibinfo{journal}{Nature} \textbf{\bibinfo{volume}{453}},
  \bibinfo{pages}{1224 } (\bibinfo{year}{2008}).

\bibitem[{\citenamefont{Yates et~al.}(2008)\citenamefont{Yates, Cohen, Ren,
  Yang, Lu, Dong, and Zhao}}]{PCT_YatesKA2008cond}
\bibinfo{author}{\bibfnamefont{K.~A.} \bibnamefont{Yates}},
  \bibinfo{author}{\bibfnamefont{L.~F.} \bibnamefont{Cohen}},
  \bibinfo{author}{\bibfnamefont{Z.-A.} \bibnamefont{Ren}},
  \bibinfo{author}{\bibfnamefont{J.}~\bibnamefont{Yang}},
  \bibinfo{author}{\bibfnamefont{W.}~\bibnamefont{Lu}},
  \bibinfo{author}{\bibfnamefont{X.-L.} \bibnamefont{Dong}}, \bibnamefont{and}
  \bibinfo{author}{\bibfnamefont{Z.-X.} \bibnamefont{Zhao}},
  \bibinfo{journal}{Supercond.\ Sci.\ Technol.} \textbf{\bibinfo{volume}{21}}
  (\bibinfo{year}{2008}).

\bibitem[{\citenamefont{Matano et~al.}()\citenamefont{Matano, Ren, Dong, Sun,
  Zhao, and qing Zheng}}]{NMR_ZhengGQ2008cond}
\bibinfo{author}{\bibfnamefont{K.}~\bibnamefont{Matano}},
  \bibinfo{author}{\bibfnamefont{Z.~A.} \bibnamefont{Ren}},
  \bibinfo{author}{\bibfnamefont{X.~L.} \bibnamefont{Dong}},
  \bibinfo{author}{\bibfnamefont{L.~L.} \bibnamefont{Sun}},
  \bibinfo{author}{\bibfnamefont{Z.~X.} \bibnamefont{Zhao}}, \bibnamefont{and}
  \bibinfo{author}{\bibfnamefont{G.}~\bibnamefont{qing Zheng}},
  \bibinfo{howpublished}{arXiv:0806.0249 [cond-mat.supr-con]}.

\bibitem[{\citenamefont{Li and Wang}(2008)}]{TwobandSc_LiJ2008CPL}
\bibinfo{author}{\bibfnamefont{J.}~\bibnamefont{Li}} \bibnamefont{and}
  \bibinfo{author}{\bibfnamefont{Y.~P.} \bibnamefont{Wang}},
  \bibinfo{journal}{Chin.\ Phys.\ Lett.} \textbf{\bibinfo{volume}{25}},
  \bibinfo{pages}{2232} (\bibinfo{year}{2008}).

\bibitem[{\citenamefont{Marsiglio and Hirsch}()}]{TwobandSc_Marsiglio2008cond}
\bibinfo{author}{\bibfnamefont{F.}~\bibnamefont{Marsiglio}} \bibnamefont{and}
  \bibinfo{author}{\bibfnamefont{J.~E.} \bibnamefont{Hirsch}},
  \bibinfo{howpublished}{arXiv:0804.0002 [cond-mat.supr-con]}.

\bibitem[{\citenamefont{Wang et~al.}()\citenamefont{Wang, Tang, Fang, and
  Dai}}]{TwobandSc_WangZH2008cond}
\bibinfo{author}{\bibfnamefont{Z.-H.} \bibnamefont{Wang}},
  \bibinfo{author}{\bibfnamefont{H.}~\bibnamefont{Tang}},
  \bibinfo{author}{\bibfnamefont{Z.}~\bibnamefont{Fang}}, \bibnamefont{and}
  \bibinfo{author}{\bibfnamefont{X.}~\bibnamefont{Dai}},
  \bibinfo{howpublished}{arXiv:0805.0736 [cond-mat.supr-con]}.

\bibitem[{\citenamefont{Han et~al.}(2008)\citenamefont{Han, Chen, and
  Wang}}]{TwobandSc_HanQ2008EPL}
\bibinfo{author}{\bibfnamefont{Q.}~\bibnamefont{Han}},
  \bibinfo{author}{\bibfnamefont{Y.}~\bibnamefont{Chen}}, \bibnamefont{and}
  \bibinfo{author}{\bibfnamefont{Z.~D.} \bibnamefont{Wang}},
  \bibinfo{journal}{EuroPhys.\ Lett.} \textbf{\bibinfo{volume}{82}},
  \bibinfo{pages}{37007} (\bibinfo{year}{2008}).

\bibitem[{\citenamefont{Weyeneth et~al.}()\citenamefont{Weyeneth, Mosele,
  Kohout, Roos, Keller, Bukowski, and Karpinski}}]{MagAnis_WeyenethS2008cond}
\bibinfo{author}{\bibfnamefont{S.}~\bibnamefont{Weyeneth}},
  \bibinfo{author}{\bibfnamefont{U.}~\bibnamefont{Mosele}},
  \bibinfo{author}{\bibfnamefont{S.}~\bibnamefont{Kohout}},
  \bibinfo{author}{\bibfnamefont{J.}~\bibnamefont{Roos}},
  \bibinfo{author}{\bibfnamefont{H.}~\bibnamefont{Keller}},
  \bibinfo{author}{\bibfnamefont{N.~D. Z. S. K.~Z.} \bibnamefont{Bukowski}},
  \bibnamefont{and}
  \bibinfo{author}{\bibfnamefont{J.}~\bibnamefont{Karpinski}},
  \bibinfo{howpublished}{arXiv:0806.1024 [cond-mat.supr-con]}.

\bibitem[{\citenamefont{Hunte et~al.}(2008)\citenamefont{Hunte, Jaroszynski,
  Gurevich, Larbalestier, Jin, Sefat, McGuire, Sales, Christen, and
  Mandrus}}]{HighMag_HunteF2008Nature}
\bibinfo{author}{\bibfnamefont{F.}~\bibnamefont{Hunte}},
  \bibinfo{author}{\bibfnamefont{J.}~\bibnamefont{Jaroszynski}},
  \bibinfo{author}{\bibfnamefont{A.}~\bibnamefont{Gurevich}},
  \bibinfo{author}{\bibfnamefont{D.~C.} \bibnamefont{Larbalestier}},
  \bibinfo{author}{\bibfnamefont{R.}~\bibnamefont{Jin}},
  \bibinfo{author}{\bibfnamefont{A.~S.} \bibnamefont{Sefat}},
  \bibinfo{author}{\bibfnamefont{M.~A.} \bibnamefont{McGuire}},
  \bibinfo{author}{\bibfnamefont{B.~C.} \bibnamefont{Sales}},
  \bibinfo{author}{\bibfnamefont{D.~K.} \bibnamefont{Christen}},
  \bibnamefont{and} \bibinfo{author}{\bibfnamefont{D.}~\bibnamefont{Mandrus}},
  \bibinfo{journal}{Nature} \textbf{\bibinfo{volume}{453}}, \bibinfo{pages}{903
  } (\bibinfo{year}{2008}).

\bibitem[{\citenamefont{Ren et~al.}(2008{\natexlab{b}})\citenamefont{Ren, Yang,
  Lu, Yi, Shen, Li, Che, Dong, Sun, Zhou et~al.}}]{HP_RenZA2008cond}
\bibinfo{author}{\bibfnamefont{Z.-A.} \bibnamefont{Ren}},
  \bibinfo{author}{\bibfnamefont{J.}~\bibnamefont{Yang}},
  \bibinfo{author}{\bibfnamefont{W.}~\bibnamefont{Lu}},
  \bibinfo{author}{\bibfnamefont{W.}~\bibnamefont{Yi}},
  \bibinfo{author}{\bibfnamefont{X.-L.} \bibnamefont{Shen}},
  \bibinfo{author}{\bibfnamefont{Z.-C.} \bibnamefont{Li}},
  \bibinfo{author}{\bibfnamefont{G.-C.} \bibnamefont{Che}},
  \bibinfo{author}{\bibfnamefont{X.-L.} \bibnamefont{Dong}},
  \bibinfo{author}{\bibfnamefont{L.-L.} \bibnamefont{Sun}},
  \bibinfo{author}{\bibfnamefont{F.}~\bibnamefont{Zhou}}, \bibnamefont{et~al.},
  \bibinfo{journal}{EuroPhys.\ Lett.} \textbf{\bibinfo{volume}{82}}
  (\bibinfo{year}{2008}{\natexlab{b}}).

\bibitem[{\citenamefont{Shan et~al.}(2003)\citenamefont{Shan, Tao, Gao, Li,
  Ren, Che, and Wen}}]{tip-prepare_ShanL2003PRB}
\bibinfo{author}{\bibfnamefont{L.}~\bibnamefont{Shan}},
  \bibinfo{author}{\bibfnamefont{H.~J.} \bibnamefont{Tao}},
  \bibinfo{author}{\bibfnamefont{H.}~\bibnamefont{Gao}},
  \bibinfo{author}{\bibfnamefont{Z.~Z.} \bibnamefont{Li}},
  \bibinfo{author}{\bibfnamefont{Z.~A.} \bibnamefont{Ren}},
  \bibinfo{author}{\bibfnamefont{G.~C.} \bibnamefont{Che}}, \bibnamefont{and}
  \bibinfo{author}{\bibfnamefont{H.~H.} \bibnamefont{Wen}},
  \bibinfo{journal}{Phys.\ Rev.\ B} \textbf{\bibinfo{volume}{68}},
  \bibinfo{pages}{144510} (\bibinfo{year}{2003}).

\bibitem[{\citenamefont{Deutscher}(2005)}]{Deutscher2005RMP}
\bibinfo{author}{\bibfnamefont{G.}~\bibnamefont{Deutscher}},
  \bibinfo{journal}{Rev.\ Mod.\ Phys.} \textbf{\bibinfo{volume}{77}},
  \bibinfo{pages}{109} (\bibinfo{year}{2005}), \bibinfo{note}{and references
  therein}.

\bibitem[{\citenamefont{Samuely et~al.}()\citenamefont{Samuely, Szab{\'o},
  Pribulov{\'a}, Tillman, Bud'ko, and Canfield}}]{PCAR_Samuely2008cond}
\bibinfo{author}{\bibfnamefont{P.}~\bibnamefont{Samuely}},
  \bibinfo{author}{\bibfnamefont{P.}~\bibnamefont{Szab{\'o}}},
  \bibinfo{author}{\bibfnamefont{Z.}~\bibnamefont{Pribulov{\'a}}},
  \bibinfo{author}{\bibfnamefont{M.~E.} \bibnamefont{Tillman}},
  \bibinfo{author}{\bibfnamefont{S.}~\bibnamefont{Bud'ko}}, \bibnamefont{and}
  \bibinfo{author}{\bibfnamefont{P.~C.} \bibnamefont{Canfield}},
  \bibinfo{howpublished}{arXiv:0806.1672 [cond-mat.supr-con]}.

\bibitem[{\citenamefont{Blonder et~al.}(1982)\citenamefont{Blonder, Tinkham,
  and Klapwijk}}]{BTK_BlonderGE1982PRB}
\bibinfo{author}{\bibfnamefont{G.~E.} \bibnamefont{Blonder}},
  \bibinfo{author}{\bibfnamefont{M.}~\bibnamefont{Tinkham}}, \bibnamefont{and}
  \bibinfo{author}{\bibfnamefont{T.~M.} \bibnamefont{Klapwijk}},
  \bibinfo{journal}{Phys.\ Rev.\ B} \textbf{\bibinfo{volume}{25}},
  \bibinfo{pages}{4515} (\bibinfo{year}{1982}).

\bibitem[{\citenamefont{Kashiwaya and Tanaka}(2000)}]{ZBCP-KashiwayaS2000RPP}
\bibinfo{author}{\bibfnamefont{S.}~\bibnamefont{Kashiwaya}} \bibnamefont{and}
  \bibinfo{author}{\bibfnamefont{Y.}~\bibnamefont{Tanaka}},
  \bibinfo{journal}{Rep.\ Prog.\ Phys.} \textbf{\bibinfo{volume}{63}},
  \bibinfo{pages}{1641} (\bibinfo{year}{2000}).

\bibitem[{\citenamefont{Aprili et~al.}(1998)\citenamefont{Aprili, Covington,
  Paraoanu, Niedermeier, and Greene}}]{ZBCP-ApriliM1998PRBr}
\bibinfo{author}{\bibfnamefont{M.}~\bibnamefont{Aprili}},
  \bibinfo{author}{\bibfnamefont{M.}~\bibnamefont{Covington}},
  \bibinfo{author}{\bibfnamefont{E.}~\bibnamefont{Paraoanu}},
  \bibinfo{author}{\bibfnamefont{B.}~\bibnamefont{Niedermeier}},
  \bibnamefont{and} \bibinfo{author}{\bibfnamefont{L.~H.}
  \bibnamefont{Greene}}, \bibinfo{journal}{Phys.\ Rev.\ B}
  \textbf{\bibinfo{volume}{57}}, \bibinfo{pages}{R8139} (\bibinfo{year}{1998}).

\bibitem[{\citenamefont{Asano and Tanaka}(2002)}]{ZBCP_Tanaka2002PRB}
\bibinfo{author}{\bibfnamefont{Y.}~\bibnamefont{Asano}} \bibnamefont{and}
  \bibinfo{author}{\bibfnamefont{Y.}~\bibnamefont{Tanaka}},
  \bibinfo{journal}{Phys.\ Rev.\ B} \textbf{\bibinfo{volume}{65}},
  \bibinfo{pages}{064522} (\bibinfo{year}{2002}).

\bibitem[{\citenamefont{Plecen\'{i}k et~al.}(1994)\citenamefont{Plecen\'{i}k,
  Grajcar, \v{S}. Be\v{n}a\v{c}ka, Seidel, and Pfuch}}]{Broad_PlecenA1994PRB}
\bibinfo{author}{\bibfnamefont{A.}~\bibnamefont{Plecen\'{i}k}},
  \bibinfo{author}{\bibfnamefont{M.}~\bibnamefont{Grajcar}},
  \bibinfo{author}{\bibnamefont{\v{S}. Be\v{n}a\v{c}ka}},
  \bibinfo{author}{\bibfnamefont{P.}~\bibnamefont{Seidel}}, \bibnamefont{and}
  \bibinfo{author}{\bibfnamefont{A.}~\bibnamefont{Pfuch}},
  \bibinfo{journal}{Phys.\ Rev.\ B} \textbf{\bibinfo{volume}{49}},
  \bibinfo{pages}{10016} (\bibinfo{year}{1994}).

\bibitem[{\citenamefont{Sachdev}(2002)}]{dxy_Sachdev2002PhysicaA}
\bibinfo{author}{\bibfnamefont{S.}~\bibnamefont{Sachdev}},
  \bibinfo{journal}{Physica A} \textbf{\bibinfo{volume}{313}},
  \bibinfo{pages}{252} (\bibinfo{year}{2002}).

\bibitem[{\citenamefont{Dagan et~al.}(2000)\citenamefont{Dagan, Krupke, and
  Deutscher}}]{Gap-DaganY2000PRB}
\bibinfo{author}{\bibfnamefont{Y.}~\bibnamefont{Dagan}},
  \bibinfo{author}{\bibfnamefont{R.}~\bibnamefont{Krupke}}, \bibnamefont{and}
  \bibinfo{author}{\bibfnamefont{G.}~\bibnamefont{Deutscher}},
  \bibinfo{journal}{Phys.\ Rev.\ B} \textbf{\bibinfo{volume}{62}},
  \bibinfo{pages}{146} (\bibinfo{year}{2000}).

\bibitem[{\citenamefont{Millo et~al.}()\citenamefont{Millo, Asulin, Yuli,
  Felner, Ren, Shen, Che, and Zhao}}]{STS_Millo2008cond}
\bibinfo{author}{\bibfnamefont{O.}~\bibnamefont{Millo}},
  \bibinfo{author}{\bibfnamefont{I.}~\bibnamefont{Asulin}},
  \bibinfo{author}{\bibfnamefont{O.}~\bibnamefont{Yuli}},
  \bibinfo{author}{\bibfnamefont{I.}~\bibnamefont{Felner}},
  \bibinfo{author}{\bibfnamefont{Z.-A.} \bibnamefont{Ren}},
  \bibinfo{author}{\bibfnamefont{X.-L.} \bibnamefont{Shen}},
  \bibinfo{author}{\bibfnamefont{G.-C.} \bibnamefont{Che}}, \bibnamefont{and}
  \bibinfo{author}{\bibfnamefont{Z.-X.} \bibnamefont{Zhao}},
  \bibinfo{howpublished}{arXiv:0807.0359 [cond-mat.supr-con]}.

\end{thebibliography}
\end{document}